\documentclass[10pt]{article}
\usepackage{fullpage,graphicx,subfigure,mathdots,mathpazo,color}
\usepackage{amsmath,amscd,tikz,mathrsfs,cite}
\usepackage[normalem]{ulem}
\usepackage{amsmath}
\usepackage{setspace}

\usepackage{epsfig,amsmath,graphicx,amssymb,overpic}
\usepackage{bm}
\usepackage{graphicx}
\usepackage{subfigure, xcolor}
\usepackage{ntheorem}

\usepackage{geometry}
\usepackage{adjustbox}

\def\be{\begin{equation}}
\def\ee{\end{equation}}
\def\bee{\begin{eqnarray}}
\def\ene{\end{eqnarray}}
\def\bes{\begin{subequations}}
\def\ees{\end{subequations}}
\def\no{\nonumber}

\newtheorem{proposition}{Proposition}

\def\det{{\rm det}}
\def\d{\displaystyle}

\def\v{\vspace{0.1in}}
\def\no{{\nonumber}}

\allowdisplaybreaks[4]

\usepackage{diagbox}
\begin{document}
\baselineskip=13pt
\renewcommand {\thefootnote}{\dag}
\renewcommand {\thefootnote}{\ddag}
\renewcommand {\thefootnote}{ }

\pagestyle{plain}

\begin{center}
\baselineskip=16pt \leftline{} \vspace{-.3in} {\Large \bf Interactions of fractional $N$-solitons with anomalous dispersions for the integrable combined fractional higher-order mKdV hierarchy} \\[0.2in]
\end{center}


\begin{center}
Minghe Zhang$^{a,b}$,\,\, Weifang Weng$^{a,b}$,\,\, and Zhenya Yan$^{a,b,*}$\footnote{$^{*}${\it Email address}: zyyan@mmrc.iss.ac.cn (Corresponding author)}  \\[0.1in]
{\it \small $^a$KLMM, Academy of Mathematics and Systems Science,  Chinese Academy of Sciences, Beijing 100190, China \\
$^b$School of Mathematical Sciences, University of Chinese Academy of Sciences, Beijing 100049, China
} \\
\end{center}

\vspace{0.1in}

\baselineskip=15pt



\noindent{\bf Abstract:}
In this paper, we investigate the anomalous dispersive relations, inverse scattering transform with a Riemann-Hilbert (RH) problem, and fractional multi-solitons of the integrable combined fractional higher-order mKdV (fhmKdV) hierarchy, including the fractional mKdV (fmKdV), fractional fifth-order mKdV (f5mKdV), fractional combined third-fifth-order mKdV (f35mKdV) equations, etc., which can be featured via completeness of squared scalar eigenfunctions of the ZS spectral problem.  We construct a matrix RH problem to present three types of fractional $N$-solitons illustrating anomalous dispersions of the combined fhmKdV hierarchy for the reflectionless case. As some examples, we analyze the wave velocity of the fractional one-soliton  such that we find that the fhmKdV equation predicts a power law relationship between the wave velocity and amplitude, and
demonstrates the anomalous dispersion. Furthermore, we illustrate other interesting anomalous dispersive wave phenomena containing the elastic interactions of fractional bright and dark solitons, W-shaped soliton and dark soliton, as well as breather and dark soliton. These obtained fractional multi-solitons will be useful to understand the related nonlinear super-dispersive wave propagations in fractional nonlinear media. \vspace{0.05in}

\vspace{0.1in} \noindent {\it Keywords:} Combined fractional mKdV hierarchy; inverse scattering; Riemann-Hilbert problem; fractional $N$-soliton solutions; anomalous dispersive relations


\vspace{0.1in}

\baselineskip=15pt

\section{Introduction}



\noindent Fractional (order) calculus (FC) almost has the same long history as the integer order calculus~\cite{fc-book,fc0}, however due to the lack of real world applications, the development of fractional calculus became very slowly~\cite{fc1,fc2}.
In fact, many complex phenomena in nature with anomalous dynamics cannot be depicted by he models with only integer order derivatives. In other words, the fractional order models were shown to be more realistic than the usual integer order models to describe some phenomena with some power-law physical quantities, such as time-dependent displacement $t^{\alpha},\,\alpha>0$~\cite{frw,frw2}, power-law potential $\phi^p,\, p>0$ for the inflaton field $\phi$~\cite{plp}. Particular attention was paid to FC up to 1970's~\cite{fc1,fc2}. Since then, it has been applied in many different areas of mathematics and physics, such as  random walks~\cite{frw,frw2}, random diffusion~\cite{31},  telomere motion~\cite{prl09}, diffusion waves~\cite{32,33,35}, turbulence~\cite{27,28,30}, control and robotics~\cite{41}, viscoelastic dynamics~\cite{42}, and finance~\cite{43,44}.

Integrable integer-order nonlinear equations are also an important class of nonlinear wave equations in the study of nonlinear dynamics, such as the Korteweg-de Vries (KdV) equation, modified Korteweg-de Vriese (mKdV) equation, Boussinesq equation, Kadomtsev-Petviashvili equation, nonlinear Schr\"odinger (NLS) equation, Hirota equation, and sine/sinh-Gordon equation~\cite{soliton,soliton-2,soliton-3}, which can be solved by the inverse scattering transform (IST)~\cite{GGKM}. It is a natural idea to extend these integer-order nonlinear equations to the fractional order cases. Many fractional physical models have been paid attention to this extension, such as the fractional NLS equation, and fractional KdV equation, and so on~\cite{carr18,boris20,boris21}, but these fractional models cannot usually guarantee the important IST integrability~\cite{fc-book2}. More recently, based on the definition of Riesz fractional derivative $|\!\!-\!\partial^2_x|^{\epsilon},\, \epsilon\in(0, 1)$~\cite{riesz,lischke}, Ablowitz and his collaborators~\cite{fmark, ab22} presented some new integrable fractional nonlinear wave equations, such as the fractional KdV, fractional NLS, fractional mKdV, and fractional sine/sinh-Gordon equations, and the powerful IST with
Gel'fand-Levitan-Marchenko (GLM) integral equation was used to find their fractional one-soliton solutions.

The Riemann-Hilbert (RH) approach~\cite{novikov} has been used to solve some integrable integer-order nonlinear wave equations~\cite{rh1,rh2,rh3,rh5,rh6,rh7,fan16,yang09,guo12}. However, to the best of our knowledge, the RH approach was not extended to solve the integrable fractional nonlinear wave equations before. In this paper, we would like to give the combined fractional higher-order mKdV (fhmKdV) hierarchy, and use the IST with RH approach to study the interactions
of the fractional multi-solitons with anomalous dispersion of the combined fhmKdV hierarchy.

The rest of this paper is organized as follows. In Sec. 2, we study the power-law dispersion relations of the combined fractional  higher-order mKdV (fhmKdV) hierarchy, and give its explicit form using the completeness relation of squared scalar eigenfunctions. In Sec. 3, with the aid of the Zakharov–Shabat spectral problem with the real-valued sufficient decay and smoothness potential and the asymptotics depending on time evolutions of eigenfunctions, the matrix RH problem is established, whose solutions are related to the solutions of the the combined fhmKdV hierarchy. By using Plemelj’s formula to solve the RH problem, we can find the fractional solutions of the combined fhmKdV hierarchy. In Sec. 4, for the reflectionless case, we present the explicitly fractional multi-solitons for the  combined fhmKdV hierarchy. In particular, we display the wave velocity of the fractional one-soliton such that we find that the fhmKdV equation predicts a power-law relation between the wave velocity and amplitude and displays the anomalous dispersion. Moreover, we illustrate other interesting super-dispersive wave phenomena containing
the elastic interactions of bright and dark solitons, W-shaped soliton and dark soliton, breather and dark soliton, as well as
bright-dark-dark solitons. Finally, we summarize our work and give some discussions in Sec. 5.

\section{Combined fractional $(2n\!+\!1)$th-order mKdV hierarchy with anomalous dispersion}

Starting from the $2\times 2$ Zakharov–Shabat spectral (eigenvalue) problem~\cite{nls}
\begin{equation}\label{laxx}
\Phi_x=X(x,t; \lambda)\Phi, \qquad X(x,t; \lambda)=i\lambda\sigma_3+U,
\end{equation}
where $\Phi=\Phi(x,t; \lambda)$ is the $2\times 2$ matrix-valued eigenfunction, $\lambda\in\mathbb{C}$ is a spectral parameter, $\sigma_3$ and the real-valued potential matrix $U(x,t)$ are defined as
\begin{equation*}
\sigma_3=\begin{pmatrix}
1& 0\\
0&-1
\end{pmatrix},\qquad
U=\begin{pmatrix}
0&u(x,t)\\
-u(x,t) & 0
\end{pmatrix},
\end{equation*}
one can, by constructing the proper time part of matrix eigenfunction, deduce the real-valued scalar combined mKdV hierarchy~\cite{wadati73,akns} containing the mKdV equation, combined third-fifth-order mKdV equations, and higher-order equations. The combined $(2n+1)$th-order mKdV hierarchy is given by~\cite{hmkdv1,hmkdv2}
\bee\label{n-mKdV}
 u_t+\sum_{\ell=1}^n\alpha_{2\ell+1}\partial_xK_{2\ell+1}[u(x,t)]=0, \quad (x,t)\in \mathbb{R}^2,\quad \alpha_{2\ell+1}\in\mathbb{R},
 \ene
where $K_{2\ell+1}[u(x,t)]$'s are defined as
\begin{align}
&K_3[u(x,t)]= u_{xx}+3u^3, \no \v\\
&K_5[u(x,t)]=u_{4x}+10(u^2u_{xx}+uu_x^2)+6u^5, \no \v\\
&K_7[u(x,t)]=u_{6x}+14(u^2u_{4x}+4uu_xu_{xxx}+3uu_{xx}^2+5u_x^2u_{xx}+5u^4u_{xx}+10u^3u_x^2)+20u^7,\no \v\\
&K_9[u(x,t)]=u_{8x}+18u^2u_{6x}+108uu_xu_{5x}+228uu_{xx}u_{4x}+210u_x^2u_{4x}+126u^4u_{xx}+138uu_{xxx}^2 \label{n-mKdV-c} \v\\
&\qquad\qquad\qquad\quad\,\, +\, 756u_xu_{xx}u_{xxx}+1008u^3u_xu_{xxx}+182u_{xx}^3+756u^3u_{xx}^2+3108u^2u_x^2u_{xx} \no\v\\
&\qquad\qquad\qquad\quad\,\, +\, 420u^6u_{xx}+798uu_x^4+1260u^5u_x^2+70u^9, \cdots. \no 
\end{align}

 Similarly, we firstly would like to study the combined fractional third-fifth-order mKdV (f35mKdV) equation
\bee \label{mkdvh}
 u_t+{\cal M}(\widehat{L})u_x=0,\quad\,\, {\cal M}(\widehat{L})
 =\big(\alpha_5 \widehat{L}-\alpha_3\big)\widehat{L}|\widehat{L}|^{\epsilon},
\ene
where $\epsilon\in (0, 1),$\, $\alpha_3,\,\alpha_5\in\mathbb{R}$ and $\widehat{L}=-\partial^2-4u^2-4u_x\partial_{-}^{-1}u$ with $\partial=\partial/\partial x,\, \partial_{-}^{-1}=\int_{-\infty}^xdy$ is called the recursion operator.  Eq.~(\ref{mkdvh}) can be rewritten as
\bee\label{fmkdv}
 u_t+|\widehat{L}|^{\epsilon}\big[\alpha_3(u_{xxx}+6u^2u_x)+\alpha_5(u_{xxxxx}+10(u^2u_{xx}+uu_x^2)_x+30u^4u_x)\big]=0.
\ene
In particular, when $\alpha_3=1,\, \alpha_5=0$, one can refind the known fmKdV equation
\bee\label{m2}
  u_t+|\widehat{L}|^{\epsilon}(u_{xxx}+6u^2u_x)=0.
\ene
As $\alpha_3=0,\, \alpha_5=1$, one has the fractional fifth-order mKdV (f5mKdV) equation
\bee\label{fmkdv5}
  u_t+|\widehat{L}|^{\epsilon}\big[u_{xxxxx}+10(u^2u_{xx}+uu_x^2)_x+30u^4u_x\big]=0.
\ene

{\it Anomalous dispersion relation}.---The formal plane wave $u(x,t)\propto e^{i[\lambda x-w(\lambda)t]}$, where
$\lambda$ is the wave number, and $w(\lambda)$  angular frequency, is employed to the  associated linearization of Eq.~(\ref{mkdvh}) to generate the dispersive relation of the linear f35mKdV equation
\bee
     w(\lambda)=\lambda{\cal M}(\lambda^2),
\ene
where the phase velocity of the wave is ${\cal M}(\lambda^2)$. We further consider the linearization of the f35mKdV equation (\ref{m2})
\bee \no
u_t+|\!\!-\!\partial^2|^{\epsilon}(\alpha_5 u_{xxxxx}+\alpha_3u_{xxx})=0,
\ene
where $|\!\!-\!\partial^2|^{\epsilon}$ stands for the Riesz fractional derivative. As a result the anomalous dispersion relation
is given by
\bee
 w(\lambda)=(\alpha_5\lambda^2-\alpha_3)\lambda^3|\lambda^2|^{\epsilon},
\ene
in which we have the corresponding phase velocity
 \bee \label{N}
  {\cal M}(\lambda^2)=\frac{w(\lambda)}{\lambda}=(\alpha_5\lambda^2-\alpha_3)\lambda^2|\lambda^2|^{\epsilon}.
 \ene

{\it The combined fhmKdV hierarchy and anomalous dispersion relations}.---Similarly to the combined f35mKdV equation \eqref{fmkdv}, one can also consider other combined fractional higher-order mKdV (fhmKdV) equations in the form
\bee \label{fmkdvh}
u_t+{\cal M}_h(\widehat{L})u_x=0,\quad {\cal M}_h(\widehat{L})=\Big(\sum_{\ell=1}^n\alpha_{2\ell+1}(-\widehat{L})^\ell\Big)
|\widehat{L}|^{\epsilon},\quad \alpha_{2\ell+1}\in\mathbb{R},
 \ene
which can be rewritten as
\bee\label{fmkdvh2}
 u_t+|\widehat{L}|^{\epsilon}\Big(\sum_{j=1}^n\alpha_{2j+1}\partial_xK_{2j+1}[u(x,t)]\Big)=0,
\ene
where $K_{2j+1}[u(x,t)]$'s are given by Eq.~(\ref{n-mKdV-c}). In particular, as $n=1, 2, 3,...$, we have the fmKdV equation, combined f35mKdV equation, combined fractional third-fifth-seventh-order mKdV (f357mKdV) equation, and etc.

We further use the formal plane wave $u(x,t)\propto e^{i[\lambda x-w_h(\lambda)t]}$ to study the linearization of the combined fhmKdV equation (\ref{fmkdvh2})
\bee \no
 u_t+|\!\!-\!\partial^2|^{\epsilon}\Big(\sum_{\ell=1}^n\alpha_{2\ell+1}u_{(2\ell+1)x}\Big)=0,
\ene
where $u_{jx}=\partial^ju/\partial x^j$, such that the anomalous dispersion relation is presented in the form
\bee
 w_h(\lambda)=\sum_{\ell=1}^n\alpha_{2\ell+1}(-1)^\ell\lambda^{2\ell+1}|\lambda^2|^{\epsilon},
\ene
 which further leads to the phase velocity
 \bee \label{Ng}
  {\cal M}_h(\lambda^2)=\frac{w_h(\lambda)}{\lambda}=\sum_{\ell=1}^n\alpha_{2\ell+1}(-\lambda^2)^{\ell}|\lambda^2|^{\epsilon}.
 \ene

To solve the combined f35mKdV equation (\ref{fmkdv}) and fhmKdV equation \eqref{fmkdvh2} using the IST with the RH problem, we need to consider the Zakharov–Shabat spectral problem \eqref{laxx}, and the associated time evolution of eigenfunction $\Phi$ in the form
\begin{equation}
\Phi_t=T\Phi, \quad
T(x,t; \lambda)=\begin{bmatrix}
  A(x,t; \lambda) & B(x,t; \lambda) \v\\  C(x,t; \lambda) & -A(x,t; \lambda)
\end{bmatrix},
\end{equation}
In general, the expressions of $A,\, B, \, C$ can not be presented explicitly, which is similar to other integrable
fractional nonlinear equations~\cite{ab22}. But one can consider the following asymptotic properties:
\bee
T_{\pm}:=\lim_{x\rightarrow \pm\infty}T(x,t;\lambda)=-i\lambda {\cal M}(4\lambda^2)\sigma_3
\ene
with ${\cal M}(4\lambda^2)$ given by Eq.~\eqref{N} for the combined f35mKdV equation (\ref{fmkdv}),
or
\bee
T_{\pm}:=\lim_{x\rightarrow \pm\infty}T(x,t;\lambda)=-i\lambda {\cal M}_h(4\lambda^2)\sigma_3
\ene
with ${\cal M}_h(4\lambda^2)$ given by Eq.~\eqref{Ng} for the combined fhmKdV equation (\ref{fmkdvh2}),
such that one has the Jost solutions $\Phi_{\pm}(x, t; k)$ satisfying the following boundary conditions
\begin{align}\label{Jost-asy}
\Phi_{\pm}(x, t; \lambda)\sim e^{i\lambda [x-{\cal M}(4\lambda^2)t]\sigma_3},\quad x\to\pm\infty
\end{align}
for the combined f35mKdV equation (\ref{fmkdv}), or
\begin{align}\label{Jost-asyg}
\Phi_{\pm}(x, t; \lambda)\sim e^{i\lambda [x-{\cal M}_h(4\lambda^2)t]\sigma_3},\quad x\to\pm\infty
\end{align}
for the combined fhmKdV equation (\ref{fmkdvh2}).

In what follows, we mainly use the case of the combined f35mKdV equation (\ref{fmkdv}) as an example to study the fractional multi-solitons. In fact, the case of combined fhmKdV equation (\ref{fmkdvh2}) is similar. Hence we introduce a modified matrix function  $J(x,t;\lambda):=\Phi(x,t;\lambda)e^{-i\lambda [x-{\cal M}(4\lambda^2)t]\sigma_3}$ such that $J$ satisfies the boundary conditions $J_{\pm}\rightarrow \mathbb{I}$, as $x\rightarrow \pm\infty$. It is easy to see that
$J$ satisfies the modified spectral problem
\begin{equation}\label{laxjx}
J_x-i\lambda[\sigma, J]=UJ,
\end{equation}
which yields
\bee\label{integral}
J_{\pm}(x,t;\lambda)=\mathbb{I}+\int_{\pm\infty}^xe^{i\lambda(x-y)\sigma_3} U(y,t)J_{\pm}(y,t;\lambda)e^{-i\lambda(x-y)\sigma_3}dy.
\ene

Let $\mathbb{C}^+=\{\lambda|\mathrm{Im} ~\lambda>0\},\,\, \mathbb{C}^-=\{\lambda|\mathrm{Im} ~\lambda<0\}$, and
$\Phi_{\pm}(x, t; \lambda)=(\Phi_{\pm 1},\, \Phi_{\pm 2})$ and $J_{\pm}(x, t; \lambda)=(J_{\pm 1},\, J_{\pm 2})$. Then
for the given $u(x,t)\in L^1\!\left(\mathbb{R}\right)$, the matrix-valued functions $\Phi_{\pm}$ and $J_{\pm}$ both have unique solutions in $\mathbb{R}$.  Moreover,
$J_{+1,-2},\, \Phi_{+1, -2}$ ($J_{-1, +2}$, $\Phi_{-1, +2}$) can be extended analytically to $\mathbb{C}^{+}$ ($\mathbb{C}^{-}$), and continuously to $\mathbb{C}^{+}\cup \mathbb{R}$ ($\mathbb{C}^{-}\cup \mathbb{R}$).
Since $\Phi_{\pm}(x, t; \lambda)$ are both fundamental solutions of the spectral problem, thus they have the relation
\bee\label{sr} \Phi_-(x,t;\lambda)=\Phi_+(x,t;\lambda)S(\lambda),\quad  \lambda\in\mathbb{R}, \quad S(\lambda)=\left(s_{ij}(\lambda)\right)_{2\times 2}, \quad |S(\lambda)|=1,
 \ene
which yields
\bee\label{s-m}
s_{ij}(\lambda)=(-1)^{i+1}\big|\Phi_{-j}(x, t; \lambda),\,\, \Phi_{+(3-i)}(x, t; \lambda)\big|,\quad i,j=1,2.
\ene
Similarly, the scattering coefficient $s_{11}(\lambda)$ ($s_{22}(\lambda)$) in $\lambda\in\mathbb{R}$ can be extended analytically to $\mathbb{C}^{-}$ ($\mathbb{C}^{+}$), and continuously to $\mathbb{C}^{-}\cup \mathbb{R}$ ($\mathbb{C}^{+}\cup \mathbb{R}$), whereas another two scattering coefficients $s_{12}(\lambda)$ and $s_{21}(\lambda)$ can not be analytically continued away from $\mathbb{R}$. 

It follows from the spectral problem \eqref{laxx} that
\bee
 \widehat{L}\phi_{-2}=4\lambda^2 \phi_{-2},\quad  L\phi_{+1}=4\lambda^2 \phi_{+1},
\ene
where $L=-\partial^2-4u^2-4u\partial_{+}^{-1}u_y,\,\,\, \partial_{+}^{-1}=\int_x^{\infty}dy$,\,\,\,
$\phi_{\pm 1}=\Phi_{+11}^2\pm \Phi_{+12}^2,\,\,\, \phi_{\pm 2}=\Phi_{-21}^2\pm \Phi_{-22}^2.$
According to the completeness of squared scalar eigenfunctions~\cite{kaup76,ab22}, one can use
${\cal M}_h(\widehat{L})$ to act on a sufficiently smooth and decaying scalar function $g(x)$ to  yield
\bee\label{jifen1}
{\cal M}_h(\widehat{L})g(x)=\frac{1}{\pi}\int_{\Gamma_{\infty}}d\lambda
{\cal M}_h(\lambda^2)s_{22}^{-2}(\lambda)\int_{\mathbb{R}}\phi_{-2}(x,\lambda)\phi_{+1}(y,\lambda)g(y)dy,
\ene
where $\Gamma_{\infty}=\lim_{{\mathcal R}\to \infty}\Gamma_{{\mathcal R}}$ with $\Gamma_{{\mathcal R}}$ being the semicircular contour in the upper
half plane evaluated from $\lambda=-{\mathcal R}$ to $\lambda={\mathcal R}$.

For the given ${\cal M}_h(\widehat{L})$ in Eq.~\eqref{fmkdvh}, one has the combined fhmKdV hierarchy in the form
\bee\label{fmkdv-exp}
u_t+\frac{1}{\pi}\int_{\Gamma_{\infty}}d\lambda
{\cal M}_h(\lambda^2)\int_{\mathbb{R}}P(x,y, \lambda)u_ydy=0,
\ene
where $P(x,y, \lambda)=s_{22}^{-2}(\lambda)\phi_{-2}(x,\lambda)\phi_{+1}(y,\lambda)$, that is,
\bee\label{fmkdv-exp2}
u_t+\frac{1}{\pi}\int_{\Gamma_{\infty}}d\lambda|4\lambda^2|^\epsilon
\int_{\mathbb{R}}P(x,y, \lambda)\Big(\sum_{\ell=1}^n\alpha_{2\ell+1}(-\widehat{L}(y))^\ell u_y\Big) dy=0,
\ene
where $\widehat{L}(y)=-\partial_y^2-4u^2(y,t)-4u_y(y,t)\partial_{-y}^{-1}u(y, t)$ with $\partial_{-y}^{-1}=\int_{-\infty}^yds$, that is,
\bee\label{fmkdv-exp2g}
u_t+\frac{1}{\pi}\int_{\Gamma_{\infty}}d\lambda|4\lambda^2|^\epsilon
\int_{\mathbb{R}}P(x,y, \lambda)\Big(\sum_{\ell=1}^n\alpha_{2\ell+1}\partial_yK_{2\ell+1}[u(y,t)]\Big) dy=0.
\ene

In particular, as $n=2$, we have the combined f35mKdV equation
\bee\label{f35mkdv}
u_t+\frac{1}{\pi}\int_{\Gamma_{\infty}}d\lambda
|4\lambda^2|^\epsilon\int_{\mathbb{R}}P(x,y, \lambda)\Big[\alpha_3(u_{yyy}+6u^2u_y)\qquad\qquad\qquad \no\\
 +\alpha_5\left(u_{yyyyy}+10(u^2u_{yy}+uu_y^2)_y+30u^4u_y\right)\Big]dy=0.
\ene



\section{Solutions of the Riemann-Hilbert problem}

It can be found that $J_{\pm}^{-1}(x,t; \lambda)$ satisfy the  the adjoint equation of Eq.~(\ref{laxjx})
\begin{align}
\Psi_x-i\lambda[\sigma_3, \Psi]=-\Psi U.
\end{align}
Let
\bee
\begin{array}{l}
J_{-}^{-1}=\left([J_{-}^{-1}]_1,\, [J_{-}^{-1}]_2\right)^{\rm T},\quad
J_{+}^{-1}=\left([J_{+}^{-1}]_1,\, [J_{+}^{-1}]_2\right)^{\rm T}.
\end{array}
\ene
Similarly, $[J_{-}^{-1}]_1,[J_{+}^{-1}]_2$ are analytic for $\lambda\in \mathbb{C}^{+}$, and $[J_{+}^{-1}]_1,[J_{-}^{-1}]_2$ are analytic for $\lambda\in \mathbb{C}^{-}$. Furthermore, it follows from Eq.~(\ref{sr}) that
\begin{align}\label{R}
e^{i\lambda [x-{\cal M}(4\lambda^2)t]\sigma_3}J_{-}^{-1}=R(\lambda)e^{-i\lambda [x-{\cal M}(4\lambda^2)t]\sigma_3}J_{+}^{-1},\quad R(\lambda)=S^{-1}(\lambda),
\end{align}
where the inverse scattering matrix is $R(\lambda)=(r_{ij}(\lambda))_{2\times 2}$ with $r_{11}(\lambda)=s_{22}(\lambda)$,\,
$r_{22}(\lambda)=s_{11}(\lambda)$,\, $r_{12}(\lambda)=-s_{12}(\lambda)$,\, $r_{21}(\lambda)=-s_{21}(\lambda)$.

Based on the above analysis about $J_{\pm},\,J_{\pm}^{-1}$, we define the following new matrices
\begin{align} \label{P12a}
& M_{+}(x, t;\lambda):=([J_{+}]_{1},[J_{-}]_{2})=J_{+}L_1+J_{-}L_2, \v\\
 \label{P12b}
&M_{-}(x,t;\lambda):=\left([J_{+}^{-1}]_1,\, [J_{-}^{-1}]_2\right)^{\rm T}=L_1J_{+}^{-1}+L_2J_{-}^{-1},
 \end{align}
where $L_{\ell}$ is a $2\times 2$ matrix, whose element of the $(\ell,\ell)$ position is equal to $1$ and others are equal to $0$.

Therefore, based on the properties of $J_{\pm}(x,t;\lambda)$ and $J_{\pm}^{-1}(x,t;\lambda)$, and definitions of $M_{\pm}(x,t;\lambda)$, one can construct a Riemann-Hilbert problem:

\begin{itemize}

\item{} Analytic conditions:\, $M_{\pm}(x,t;\lambda)$ are analytic for $\lambda\in\mathbb{C}^{\pm}$;

\item{} Jump condition:
\begin{equation}\label{RH}
M_{-}(x,t;\lambda)M_{+}(x,t;\lambda)=G(x,t;\lambda),\quad \lambda\in\mathbb{R},
\end{equation}
where the jump matrix is
\begin{equation}\no
G(x,t;\lambda)=e^{i\lambda [x-{\cal M}(4\lambda^2)t]\sigma_3}\left(\begin{array}{cc} 1 & s_{12} \\
r_{21} & 1 \end{array}\right)e^{-i\lambda [x-{\cal M}(4\lambda^2)t]\sigma_3};
\end{equation}

\item {} $M_{\pm}(x,t;\lambda)\rightarrow \mathbb{I}$\, as \,$\lambda\rightarrow\infty.$
\end{itemize}

In fact, $M_{+}(x,t;\lambda)$ has the following  asymptotic expansion at $\lambda\rightarrow\infty$,
\begin{align}\label{p+}
M_{+}(x,t;\lambda)=\mathbb{I}+\frac{M_{+}^{(1)}(x,t)}{\lambda}+O(\lambda^{-2}),\quad \lambda\rightarrow\infty.
\end{align}
which is substituted into Eq.~\eqref{laxjx} to yield the solutions of the combined f35mKdV equation (\ref{f35mkdv})
\begin{align}\label{qju}
u(x,t)=-2i\big(M_{+}^{(1)}(x,t)\big)_{12},
\end{align}
that is, the solution of Eq.~(\ref{fmkdv}) can be converted to solve the above-mentioned Riemann-Hilbert problem.


To present the solutions for the Riemann-Hilbert problem (\ref{RH}), we need to solve the problem under the assumption of irregular, which means that both $\det M_{+}$ and $\det M_{-}$ should be zero for some $\lambda$ and all these zeros are simple. The determinants of $M_{+}$ and $M_{-}$ can be recoverd by the elements of $S(\lambda)$ and $R(\lambda):=S^{-1}(\lambda)$:
\begin{align}\label{detP}
\det M_{+}(x,t;\lambda)=r_{11}(\lambda)=s_{22}(\lambda), \quad \lambda\in\mathbb{C}^{+},\\
\det M_{-}(x,t;\lambda)=s_{11}(\lambda), \quad \lambda\in\mathbb{C}^{-}.
\end{align}
According to $U^{\dag}=-U$, one has
$S^{\dag}(\lambda^{*})=R(\lambda),\,\,
M_{+}^{\dag}(\lambda^{*})=M_{-}(\lambda),\,\, \lambda\in\mathbb{C}^{-}.$
Therefore, 
one can find that
$\det M_{+}(\lambda)=\big(\det M_{-}(\lambda^{*})\big)^{*},\,\, \lambda\in \mathbb{C}^{+}.$
Beside, it follows from $X(\lambda)=X(-\lambda^{*})^{*}$ with $U^*=U$ that one has $S(-\lambda^{*})^{*}=S(\lambda)$, which leads to $s_{11}(-\lambda^{*})^{*}=s_{11}(\lambda)$ and $\det M_{+}(\lambda)=\big(\det M_{+}(-\lambda^{*})\big)^{*}$, $\det M_{-}(\lambda)=\big(\det M_{-}(-\lambda^{*})\big)^{*}$.
Hence, one can assume that if $\det M_{+}$ has $N_1+2N_2$ simple zeros $\lambda_{1},\dots,\lambda_{N_1}$ in $i\mathbb{R}^+$ and $\lambda_{N_1+1},\cdots,\lambda_{N_1+N_2},\,-\lambda_{N_1+1}^{*},\cdots,$\,$-\lambda_{N_1+N_2}^{*}$ in $\mathbb{C}^{+}\backslash i\mathbb{R}^+$, then $\det M_{-}$ has $N_1+2N_2$ simple zeros $\lambda_{1}^*$, $\lambda_{2}^*$, $\cdots$,$\lambda_{N_1}^*$ in $i\mathbb{R}^-$ and $\lambda_{N_1+1}^*$,\,$\cdots$,\,$\lambda_{N_1+N_2}^*$,\,$-\lambda_{N_1+1}$,\, $\cdots$, \, $-\lambda_{N_1+N_2}$ in $\mathbb{C}^{-}\backslash i\mathbb{R}^-$.

Suppose the non-zero column vector $w_j \, (j=1,2,\cdots, N_1+2N_2)$ satisfy
\begin{align}\label{Pw}
M_{+}(x,t; \lambda_j)w_{j}=0.
\end{align}
Then taking the Hermitian of Eq.~(\ref{Pw}) yields
\begin{align}\label{wP}
w_{j}^{\dag}M_{-}(x,t; \lambda^{*})=0.
\end{align}

With the above discussions about the symmetry, the discrete spectra are required to satisfy one of the following three cases, in which $w_j$ can be found by taking the derivative about $x$ and $t$ of Eqs.~(\ref{Pw})-(\ref{wP}) ($\theta_{\epsilon,j}=i\lambda_j [x-{\mathcal M}(4\lambda_j^2)t]$):

\begin{itemize}

\item {} {\it Case a.}\, $N=N_1,\, N_1\in\mathbb{N}^{+}$, $\lambda_j\in i\mathbb{R}^{+}$ for $1\leqslant j \leqslant N_1$. In this case, $w_{j}=e^{\theta_{\epsilon,j}\sigma_3}w_{j0}$, where $w_{j0}$ are the real constant column vectors.

\item {} {\it Case b.}\,  $N=2N_2$, $N_2\in\mathbb{N}^{+}$, $\lambda_{j+N_2}=-\lambda_{j}^{*}\in \mathbb{C}^{+}\backslash i\mathbb{R}^{+}$ for $1\leqslant j \leqslant N_2$. In this case,
    $w_{j+}=e^{\theta_{\epsilon,j}\sigma_3}w_{j0}$,\,\, $w_{(j+N_2)+}=w_{j+}^{*}$,  where $w_{j0}$ are the complex constant column vectors.

\item {} {\it Case c.}\,  $N=N_1+2N_2$, $N_1, N_2\in\mathbb{N}^{+}$, $\lambda_j\in i\mathbb{R}^{+}$, $\lambda_{N_1+N_2+\ell}=-\lambda_{N_1+\ell}^{*}\in \mathbb{C}^{+}\backslash i\mathbb{R}^{+}$ for $1\leqslant j \leqslant N_1,\,\, 1\leqslant \ell \leqslant N_2$. In this case, $w_{j}=e^{\theta_{\epsilon,j}\sigma_3}w_{j0},\,\,
    w_{N_1+\ell}=e^{\theta_{\epsilon,j}\sigma_3}w_{(N_1+\ell)0}$ and $w_{N_1+N_2+\ell}=w_{N_1+\ell}^*,$ where $w_{j0},\, w_{(N_1+\ell)0}$ are the  real and complex constant column vectors, respectively.

\end{itemize}

\begin{proposition} The non-regular Riemann-Hilbert problem about $M_{\pm}$ with zero eigenvalues (\ref{Pw})-(\ref{wP}) can be written as
\begin{align}\label{PP}
M_{+}(x,t;\lambda)=\widehat{M}_{+}(x,t;\lambda)E(x,t;\lambda),\\
M_{-}(x,t;\lambda)=E^{-1}(x,t;\lambda)\widehat{M}_{-}(x,t;\lambda),
\end{align}
where
\begin{align} \no
E(x,t;\lambda)=\mathbb{I}+\sum_{k,j=1}^N\frac{w_kw_j^{\dag}(\Omega^{-1})_{kj}}{\lambda-\lambda_j^{*}},\quad
E^{-1}(x,t;\lambda)=\mathbb{I}-\sum_{k,j=1}^N\frac{w_kw_j^{\dag}(\Omega^{-1})_{kj}}{\lambda-\lambda_j},
\end{align}
$\Omega$ is an $N\times N$ matrix with its $(k,j)$th element given by
\begin{align}
\Omega_{kj}=\frac{w_k^{\dag}w_j}{\lambda_k^{*}-\lambda_j},\quad 1\leqslant k,j\leqslant N,\quad
\det E(\lambda)=\prod_{k=1}^{N}\frac{\lambda-\lambda_k}{\lambda-\lambda_k^{*}},
\end{align}
and $\widehat{M}_{\pm}(x,t;\lambda)$ satisfy the following regular Riemann-Hilbert problem:
\begin{itemize}

\item{} Analytic conditions:\, $\widehat{M}_{\pm}(x,t;\lambda)$ are analytic for $\lambda\in\mathbb{C}^{\pm}$,

\item{} Jump condition: $\widehat{M}_{-}(x,t;\lambda)\widehat{M}_{+}(x,t;\lambda)=E(\lambda)G(x,t;\lambda)E^{-1}(\lambda), \quad \lambda\in\mathbb{R}$,

\item {} $\widehat{M}_{\pm}(x,t;\lambda)\rightarrow \mathbb{I},\quad \lambda\rightarrow\infty.$
\end{itemize}
\end{proposition}

Through the Plemelj's formula, we have
\bee\label{hatp}
\begin{array}{rl}
\widehat{M}_{+}^{-1}(\lambda)&\displaystyle =\mathbb{I}+\dfrac{1}{2\pi i}\int_{-\infty}^{\infty}\frac{E(s)(\mathbb{I}-G(s))E^{-1}(s)M_{+}^{-1}(s)}{s-\lambda}ds \vspace{0.1in}\\
&\displaystyle =\mathbb{I}-\dfrac{1}{2\pi i\lambda}\int_{-\infty}^{\infty}E(s)(\mathbb{I}-G(s))E^{-1}(s)M_{+}^{-1}(s)ds+O(\lambda^{-2}).
\end{array}
\ene
Similarly, one has
\begin{align}\label{H}
E(\lambda)=\mathbb{I}+\frac{1}{\lambda}\sum_{k,j=1}^{N}w_kw_j^{\dag}(\Omega^{-1})_{kj}+O(\lambda^{-2}).
\end{align}

Therefore, we have
\begin{align}\label{qjgs}
M_{+}^{(1)}(x,t)=\frac{1}{2\pi i}\int_{-\infty}^{\infty}E(\mathbb{I}-G(s))E^{-1}M_{+}^{-1}ds+\sum_{k,j=1}^{N}w_kw_j^{\dag}(\Omega^{-1})_{kj}.
\end{align}

\section{Fractional $N$-solitons with anomalous dispersions}

In particular, for the reflectionless case $G=\mathbb{I}$, Eq.~(\ref{qjgs}) becomes
\begin{align}\label{qjgs1}
M_{+}^{(1)}(x,t)=\sum_{k,j=1}^{N}w_kw_j^{\dag}(\Omega^{-1})_{kj}.
\end{align}

Let $w_{j0}=(a_{1j},\, a_{2j})^{T}$, then one has
\bee
w_j=(a_{1j}e^{\theta_{\epsilon,j}},\, a_{2j}e^{-\theta_{\epsilon,j}})^{T}, \quad \theta_{\epsilon,j}=i\lambda_j [x-(16\alpha_5\lambda_j^4-4\alpha_3\lambda_j^2)|4\lambda_j^2|^{\epsilon}t],\,\,\,\,\,
1\leqslant j\leqslant N.
\ene
Substituting $w_j$ into Eq.~(\ref{qjgs1}), and using the transform (\ref{qju}), we have the fractional $N$-soliton solutions of the
combined f35mKdV equation (\ref{fmkdv}) in the form
\begin{align}\label{u}
u^{[N]}(x,t)=-2i(M_{+}^{(1)}(x,t))_{12}
 =-2i\sum_{j,k=1}^{N}a_{1k}a_{2j}^{*}e^{\theta_{\epsilon,k}-\theta_{\epsilon,j}^*}(\Omega^{-1})_{kj}
= 2i\frac{\det\, H}{\det\, \Omega},
\end{align}
where
\begin{equation}\no
H=
\left(
\begin{array}{cccc}
0 & a_{11}e^{\theta_{\epsilon,1}} & \cdots & a_{1N}e^{\theta_{\epsilon, N}} \v\\
a_{21}^{*}e^{-\theta_{\epsilon,1}^{*}} & \Omega_{11} & \cdots & \Omega_{1N}\v\\
\vdots & \vdots & \ddots & \vdots \v\\
a_{2N}^{*}e^{-\theta_{\epsilon,N}^{*}} & \Omega_{N1} & \cdots & \Omega_{NN}
\end{array}
\right),
\end{equation}
and $\Omega=(\Omega_{kj})_{N\times N}$ with
\begin{align}\label{omega}
\Omega_{kj}=\frac{a_{1k}^{*}a_{1j}e^{\theta_{\epsilon,k}^{*}+\theta_{\epsilon,j}}
+a_{2k}^{*}a_{2j}e^{-\theta_{\epsilon,k}^{*}-\theta_{\epsilon,j}}}{\lambda_k^{*}-\lambda_j},
\quad k,\, j=1,2,\cdots, N.
\end{align}

Similarly, we can find the fractional multi-solitons of the combined fractional higher-order mKdV equation (\ref{fmkdvh2}) in the form (\ref{u}) with
\bee\label{theta-h}
\theta_{\epsilon,j}\to \theta_{\epsilon,j}^{(h)}=i\lambda_j \bigg[x-\Big(\sum_{\ell=1}^n\alpha_{2\ell+1}(-4\lambda_j^2)^{\ell}\Big)|4\lambda_j^2|^{\epsilon}t\bigg],\quad
1\leqslant \ell\leqslant n.
\ene

In what follows, we will illustrate the fractional multi-solitons \eqref{u} of the f35mKdV equations for $N=1,2,3$. In fact, one can also exhibit the fractional multi-solitons \eqref{u} with \eqref{theta-h} of the combined fhmKdV equation.

\v {\bf Case 1.}\, As $N=N_1=1$, $\lambda_1=i\eta\, (\eta\in\mathbb{R}^+)$, and $a_{11},\, a_{21}\in\mathbb{R}\!\setminus\!\{0\}$, it follows from Eq.~\eqref{u} that the fractional one-soliton solution of Eq.~(\ref{fmkdv}) can be written as
\bee\label{u1}
\begin{array}{rl}
u^{[1]}(x,t)=&\!\! \d -\frac{2ia_{11}a_{21}(\lambda_1^{*}-\lambda_{1})e^{\theta_{\epsilon,1}
-\theta_{\epsilon,1}^{*}}}{a_{11}^2e^{\theta_{\epsilon,1}+\theta_{\epsilon,1}^{*}}
+a_{21}^2e^{-\theta_{\epsilon,1}-\theta_{\epsilon,1}^{*}}} \v\\
=& \!\! \d -2\eta\,{\rm sgn}(a_{11}a_{21})\,{\rm sech}
\Big\{2\eta\!\left[x-(4\alpha_5\eta^2+\alpha_3)(2\eta)^{2+2\epsilon}t\right]+\ln|a_{21}/a_{11}|\Big\}.
 \end{array}
 \ene
In particular, when $\alpha_3=1,\,\ \alpha_5=0$ and ${\rm sgn}(a_{11}a_{21})=-1$, we refind the fractional one-soliton solution of the fmKdV equation~\cite{ab22}. As $\alpha_3=0,\,\ \alpha_5=1$, we have the fractional one-soliton solutions of the f5mKdV equation. When $\alpha_3\alpha_5\not=0$, we find the fractional one-soliton solutions of the f35mKdV equation.

\begin{figure}[!t]
    \centering
\vspace{-0.15in}
  {\scalebox{0.45}[0.42]{\includegraphics{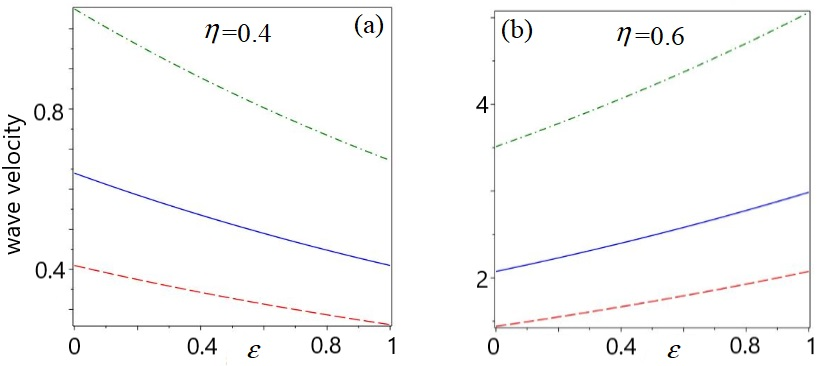}}}
\vspace{-0.05in}
\caption{\small Wave velocities \eqref{wv} vs $\epsilon$ of fractional solitons for the fmKdV equation ($\alpha_3=1$, $\alpha_5=0$, red dashed line), f5mKdV equation ($\alpha_3=0$, $\alpha_5=1$, blue dash-dotted line) and f35mKdV equation ($\alpha_3=\alpha_5=1$, green solid line) equations. (a) $\lambda_1=i\eta=0.4i$\, (i.e., $\eta=0.4)$; (b) $\lambda_1=i\eta=0.6i$\, (i.e., $\eta=0.6)$.}
  \label{fig1}
\end{figure}
\begin{figure}[!t]
    \centering
\vspace{-0.05in}
  {\scalebox{0.45}[0.43]{\includegraphics{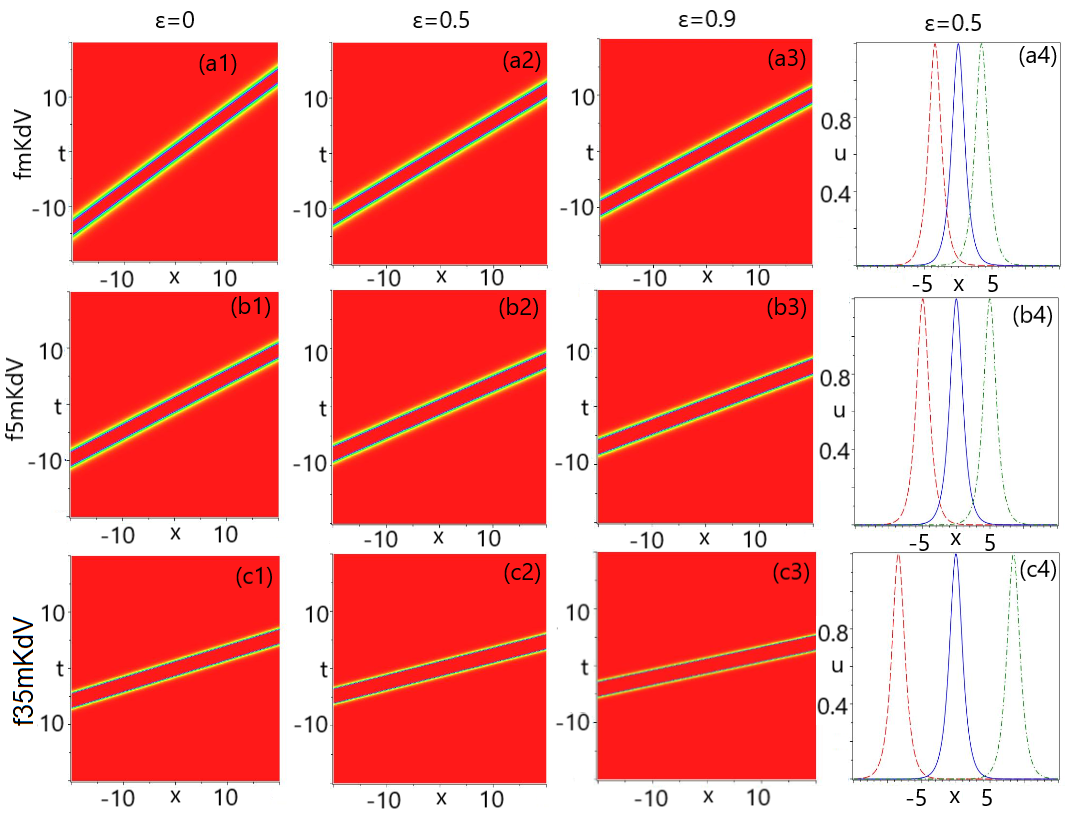}}}
\vspace{-0.05in}
\caption{\small $N=N_1=1$. Fractional one-soliton solutions \eqref{u1} for $\lambda_1=i\eta=0.6i,\,a_{11}=1,\, a_{21}=-1$. (a1)-(a4) $\alpha_3=1, \alpha_5=0$ (fmKdV equation), (b1)-(b4): $\alpha_3=0,\, \alpha_5=1$ (f5mKdV equation), (c1)-(c4): $\alpha_3=1,\, \alpha_5=1$ (f35mKdV equation). (a4)-(c4) the fractional one-soliton curves with $\epsilon=0.5$ at different times $t=-2$ (dashed line),
$t=0$ (solid line) and $t=2$ (dash-dotted line).}
  \label{fig2}
\end{figure}

The wave velocity of the fractional one-soliton solution is
\begin{equation}\label{wv}
v_{\epsilon}(\eta)=(4\alpha_5\eta^2+\alpha_3)(2\eta)^{2+2\epsilon},
\end{equation}
which depends on the parameter $\epsilon$ for the given $\alpha_{3,5}$ and $\eta$, and implies that
the f35mKdV equation predicts a power law relationship between the wave velocity $v_{\epsilon}(\eta)$ depending on $\epsilon\in(0, 1)$ and wave amplitude $\eta$, and displays the anomalous dispersion. The absolute value of the wave velocity, $|v_{\epsilon}(\eta)|$ is $|4\alpha_5\eta^2+\alpha_3|(2\eta)^{1+2\epsilon}$ times more than the amplitude $2\eta$. For the given parameter $\epsilon$, the taller solitons travel more quickly than shorter ones. Furthermore, it follows from Eq.~(\ref{wv}) that
\begin{itemize}

\item {} As $\alpha_3,\, \alpha_5\geq 0$ with $\alpha_3^2+\alpha_5^2\not=0$, one has $v_{\epsilon}(\eta)>0$, in which
 the wave is a right-going travelling wave;

\item {}  As $\alpha_3,\, \alpha_5\leq 0$ with $\alpha_3^2+\alpha_5^2\not=0$, one has $v_{\epsilon}(\eta)<0$, in which
 the wave is a left-going travelling wave;

\item {} As $\alpha_3\alpha_5<0$ and $4\alpha_5\eta^2+\alpha_3>0$, one has $v_{\epsilon}(\eta)>0$, in which
 the wave is a right-going travelling wave;

\item {}  As $\alpha_3\alpha_5<0$ and $4\alpha_5\eta^2+\alpha_3<0$, one has $v_{\epsilon}(\eta)<0$, in which
 the wave is a left-going travelling wave;

\item {} As $\alpha_3\alpha_5<0$ and $4\alpha_5\eta^2+\alpha_3=0$, one has $v_{\epsilon}(\eta)=0$, in which
 the wave is a stationary one.

\end{itemize}

\begin{figure}[!t]
    \centering
\vspace{-0.05in}
  {\scalebox{0.42}[0.4]{\includegraphics{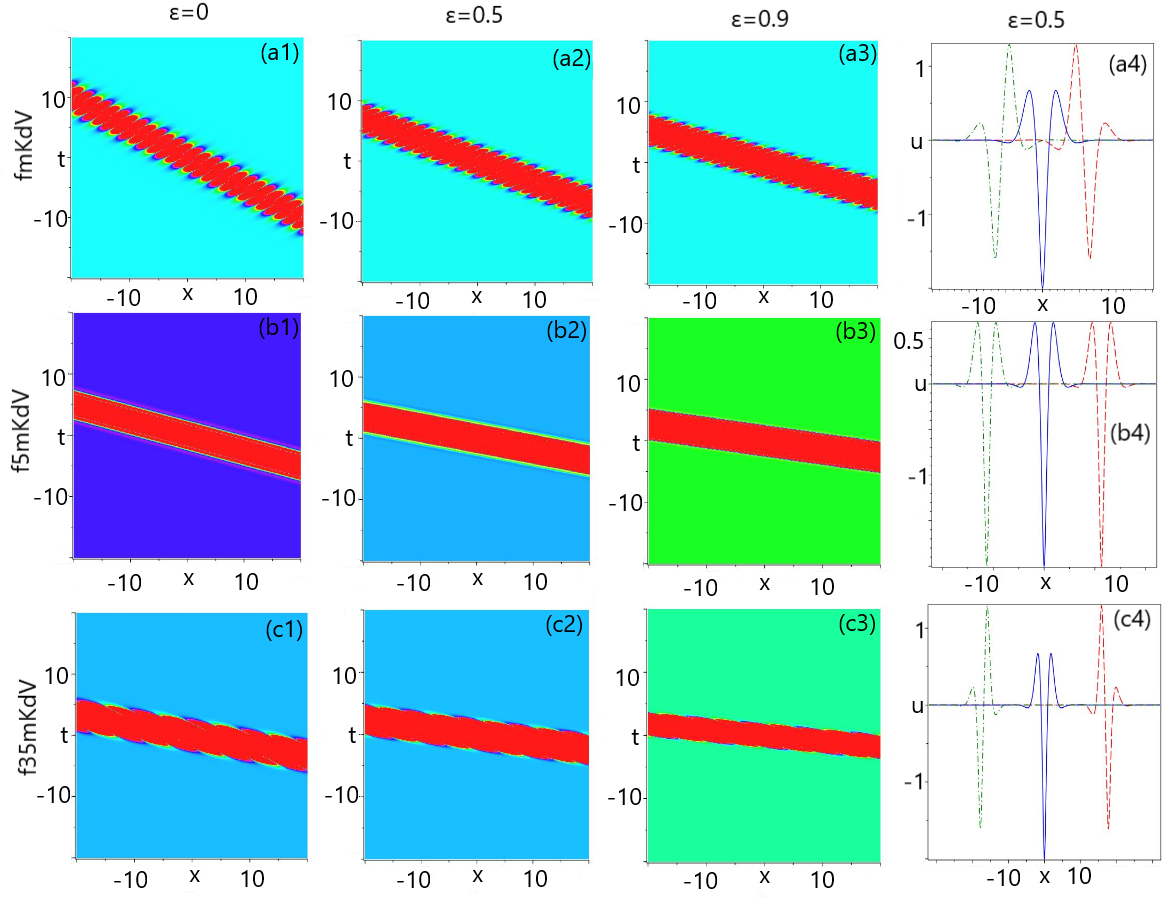}}}
\vspace{-0.05in}
\caption{\small $N=2N_2=2$. Fractional one-breather (or W-shaped soliton) solutions for $\lambda_1=-\lambda_2^*=0.5+0.5i,\,a_{11}=a_{12}=a_{21}=a_{22}=1$. (a1)-(a4) $\alpha_3=1, \alpha_5=0$ (fmKdV equation), (b1)-(b4) $\alpha_3=0,\, \alpha_5=1$ (W-shaped soliton of the f5mKdV equation), (c1)-(c4): $\alpha_3=1,\, \alpha_5=1$ (f35mKdV equation). (a4)-(c4) the fractional one-soliton curves with $\epsilon=0.5$ at different times $t=-2$ (dashed line), $t=0$ (solid line) and $t=2$ (dash-dotted line).}
  \label{fig3}
\end{figure}

Figs.~\ref{fig1}(a) and (b) display the wave velocities of the fractional one-soliton solutions for equation parameters $\alpha_3,\, \alpha_5\geq 0$ with $\alpha_3^2+\alpha_5^2\not=0$ and two spectral parameters $\lambda_1=i\eta=0.4i, \, 0.6i$, respectively. It follows from Fig.~\ref{fig1}(a) that for the given $\eta=0.4\in (0, 0.5)$, all wave velocities are decreasing functions in
$\epsilon\in [0, 1]$, however, as $\eta=0.6\in (0.5, 1)$, all wave velocities are increasing functions in
$\epsilon\in [0, 1]$ (see  Fig.~\ref{fig1}(b)). In particular, as $\eta=0.5$, one has $v_{\epsilon}=4\eta^2(4\alpha_5\eta^2+\alpha_3)$, which is independent of $\epsilon$.
It follows from Fig.~\ref{fig1}(a-b) that for the fixed $\varepsilon$, their wave velocities have the relation: $v_{\rm f35mKdV}$ $>$ $v_{\rm f5mKdV}$ $>$ $v_{\rm fmKdV}$ for $\alpha_3,\,\alpha_5\geq 0$.

Figures~\ref{fig2}(a1)-(c4) illustrate the fractional one-soliton solutions of the fmKdV, f5mKdV and f35mKdV equations for $\lambda_1=0.6i$ and different parameter $\epsilon=0,\, 0.5,\, 0.9$,  respectively, which are all right-going travelling-wave solitons without dissipating or spreading out.

\v Similarly, as $N=N_1=1$, that is, $\lambda_1=i\eta,\, (\eta\in\mathbb{R}^+)$, we can also find the fractional one-soliton solutions of the combined fractional higher-order mKdV equation (\ref{fmkdvh2}) in the form
\bee\label{u1h}
u^{[1]}(x,t)=-2\eta\,{\rm sgn}(a_{11}a_{21})\,{\rm sech}
\bigg\{2\eta\Big[x-\Big(\d\sum_{\ell=1}^n\alpha_{2\ell+1}(2\eta)^{2\ell}\Big)(2\eta)^{2\epsilon}t\Big]
+\ln|a_{21}/a_{11}|\bigg\},
 \ene
whose wave velocity is
\begin{equation}\label{wv2}
v_{h,\epsilon}(\eta)=\Big(\sum_{\ell=1}^n\alpha_{2\ell+1}(2\eta)^{2\ell}\Big)(2\eta)^{2\epsilon}.
\end{equation}
The absolute value of the wave velocity, $|v_{h,\epsilon}(\eta)|$ is $\Big|\sum_{\ell=1}^n\alpha_{2\ell+1}(2\eta)^{2\ell-1}\Big|(2\eta)^{2\epsilon}$ times more than the amplitude $2\eta$. For the given parameter $\epsilon$, the taller fractional soliton  propagates more quickly than the shorter one.

We have the following conclusions about the fractional soliton for the parameters $\eta$,\, $\alpha_{2\ell+1}$,\, $\ell=1,2,\cdots, n$, and $\epsilon$:

\begin{itemize}

\item {} When $\sum_{\ell=1}^n\alpha_{2\ell+1}(2\eta)^{2\ell}>0$, the fractional one-soliton solution \eqref{u1h} is a right-going travelling wave;

\item {} When $\sum_{\ell=1}^n\alpha_{2\ell+1}(2\eta)^{2\ell}<0$, the fractional one-soliton solution \eqref{u1h} is a left-going travelling wave;

 \item {} When $\sum_{\ell=1}^n\alpha_{2\ell+1}(2\eta)^{2\ell}=0$, the fractional one-soliton solution \eqref{u1h} is a stationary wave.

\item {} As $\sum_{\ell=1}^n\alpha_{2\ell+1}(2\eta)^{2\ell}> (<)\, 0$ and $0<\eta<0.5$, the wave velocity $v_{h,\epsilon}(\eta)> (<)\,0$ and it is a decreasing (increasing) function of $\epsilon$;

\item {} As $\sum_{\ell=1}^n\alpha_{2\ell+1}(2\eta)^{2\ell}> (<)\, 0$ and $\eta>0.5$, the wave velocity $v_{h,\epsilon}(\eta)> (<)\,0$ and it is a increasing (decreasing) function of $\epsilon$;

\item {} As $\eta=0.5$, the wave velocity a constant function of $\epsilon$, i.e. $v_{h,\epsilon}(\eta)=\sum_{\ell=1}^n\alpha_{2\ell+1}(2\eta)^{2\ell}$.
\end{itemize}

Notice that since there are the more parameters $\alpha_{2\ell+1},\, \ell=1,2,\cdots, n$ in the combined fhmKdV equation (\ref{fmkdvh2}), thus the fractional one-soliton solution can generate more wave structures than the fmKdV equation (\ref{m2}).

\begin{figure}[!t]
    \centering
\vspace{-0.15in}
  {\scalebox{0.3}[0.3]{\includegraphics{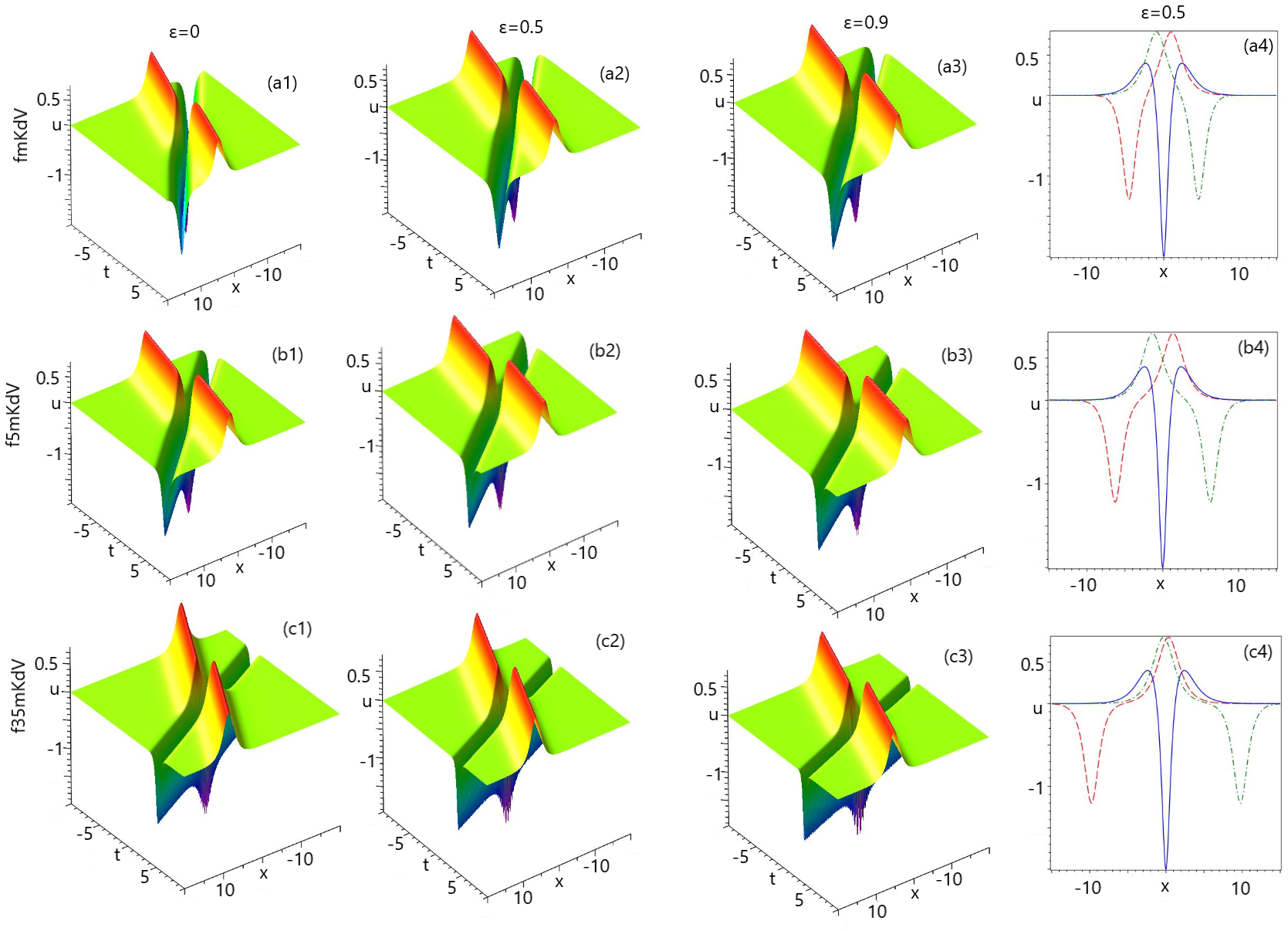}}}
\vspace{-0.1in}
\caption{\small $N=N_1=2.$ Elastic interactions of the fractional two-soliton (bright-dark solitons) solutions for $\lambda_1=0.4i,\,\lambda_2=0.6i,\,a_{11}=a_{12}=a_{21}=a_{22}=1$. (a1)-(a4) $\alpha_3=1, \alpha_5=0$ (fmKdV equation), (b1)-(b4): $\alpha_3=0,\, \alpha_5=1$ (f5mKdV equation), (c1)-(c4): $\alpha_3=1,\, \alpha_5=1$ (f35mKdV equation). (a4)-(c4) the fractional one-soliton curves with $\epsilon=0.5$ at different times $t=-2$ (dashed line), $t=0$ (solid line) and $t=2$ (dash-dotted line).}
  \label{fig2pure}
\end{figure}

\v {\bf Case 2a.}\, As $N=2N_2=2$, that is, a pair of anti conjugate spectral parameters $\lambda_1\in\mathbb{C}^{+}\backslash i\mathbb{R}^{+}$ and $\lambda_2=-\lambda_1^*$ is considered,  the fractional one-breather (or W-shaped soliton) solutions of Eq.~(\ref{fmkdv}) are given by
\bee
u^{[2]}(x,t)\!=\!\dfrac{2i(a_{11}a_{21}^{*}e^{\theta_{\epsilon,1}-\theta_{\epsilon,1}^*}\Omega_{22}
\!-\!a_{11}a_{22}^{*}e^{\theta_{\epsilon,1}-\theta_{\epsilon,2}^*}\Omega_{12}
\! -\!a_{12}a_{21}^{*}e^{\theta_{\epsilon,2}-\theta_{\epsilon,1}^*}\Omega_{21}
 \!+\! a_{12}a_{22}^{*}e^{\theta_{\epsilon,2}-\theta_{\epsilon,2}^*}\Omega_{11})}
 {\Omega_{12}\Omega_{21}-\Omega_{11}\Omega_{22}},
\ene
where $\Omega_{kj},\, k,j=1,2$ are given by Eq.~\eqref{omega}.

\begin{figure}[!t]
    \centering
\vspace{-0.15in}
  {\scalebox{0.3}[0.3]{\includegraphics{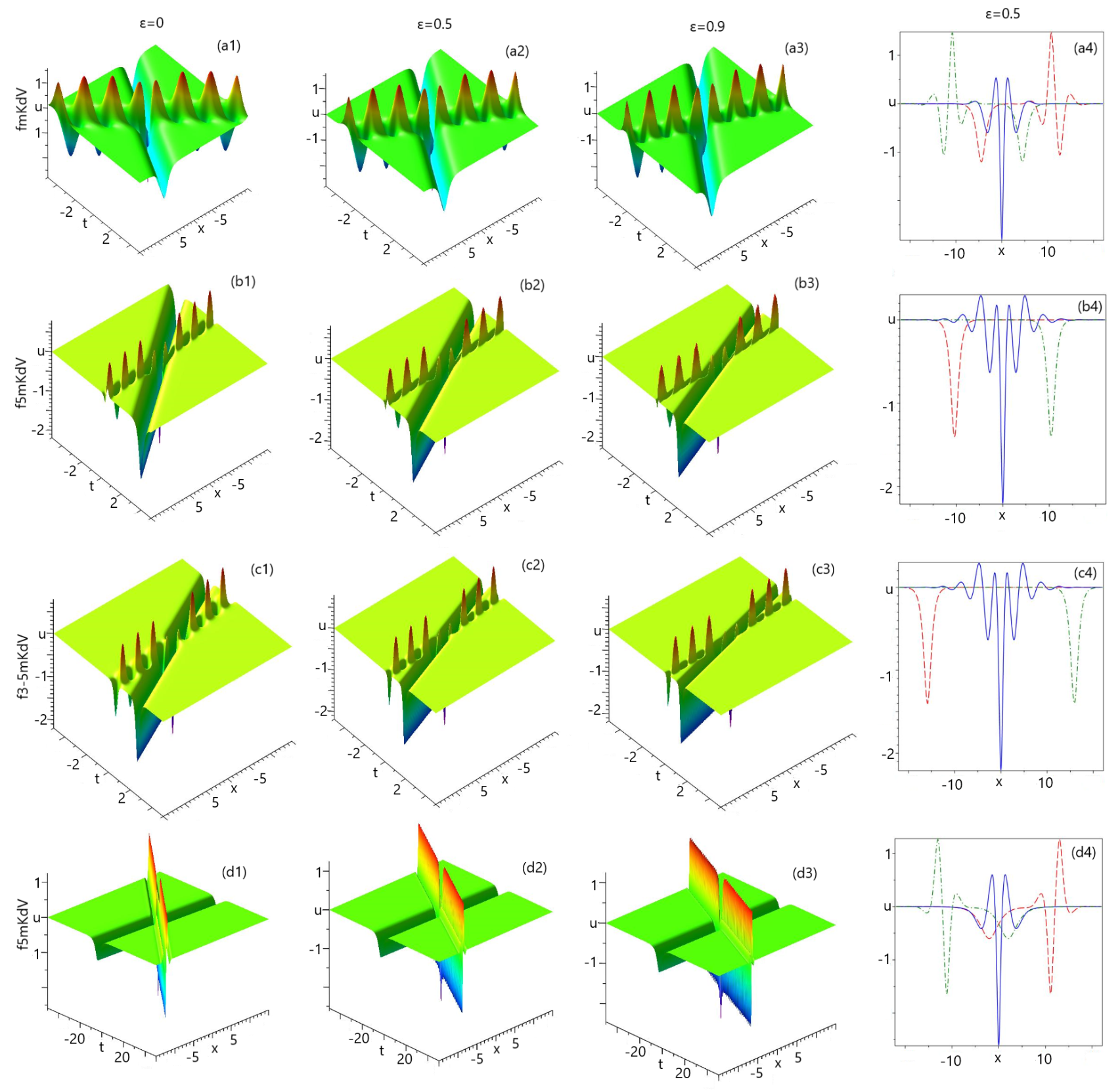}}}
\vspace{-0.1in}
\caption{\small $N=N_1+2N_2=3,\, N_1=1,\, N_2=1$ and $a_{ij}=1,\,i=1,2; j=1,2,3$.
Elastic interactions of fractional one-breather solution (or W-shaped soliton) and one-dark-soliton solution for
(a1)-(a4) $\lambda_1=0.6i,\, \lambda_2=-\lambda_3^*=0.6+0.4i,\, \alpha_3=1, \alpha_5=0$ (fmKdV equation),
(b1)-(b4) $\lambda_1=0.7i,\, \lambda_2=-\lambda_3^*=0.8+0.2i,\, \alpha_3=0,\, \alpha_5=1$ (f5mKdV),
(c1)-(c4): $\lambda_1=0.7i,\, \lambda_2=-\lambda_3^*=0.8+0.2i,\, \alpha_3=1,\, \alpha_5=1$ (f35mKdV equation);
(d1)-(d4) $\lambda_1=0.3i,\, \lambda_2=-\lambda_3^*=0.5+0.5i,\, \alpha_3=0,\, \alpha_5=1$ (f5mKdV),
(a4)-(d4) the fractional two-soliton curves with $\epsilon=0.5$ at different times $t=-2$ (dashed line), $t=0$ (solid line) and $t=2$ (dash-dotted line).}
  \label{fig4}
\end{figure}

Figures~\ref{fig3}(a1)-(a4) and (c1)-(c4) illustrate, respectively, the fractional one-breather solutions of the fmKdV
and f35mKdV equations for two spectral parameters $\lambda_1=-\lambda_2^*=0.5+0.5i$ and different parameters $\epsilon=0,\, 0.5,\, 0.9$, which are all left-going travelling-wave solitons without dissipating or spreading out. However,
Figures~\ref{fig3}(b1)-(b4) exhibit the fractional W-shaped one-soliton solutions of the f5mKdV equation for two spectral parameters $\lambda_1=-\lambda_2^*=0.5+0.5i$ and different parameters $\epsilon=0,\, 0.5,\, 0.9$.

\v {\bf Case 2b.}\, As $N=N_1=2$, that is,  two spectral parameters $\lambda_1,\, \lambda_2\in i\mathbb{R}^+$ are considered,
the fractional two-soliton (bright-dark) solutions of Eq.~(\ref{fmkdv}) are given by
\bee
u^{[2]}(x,t)\!=\!\dfrac{2i(a_{11}a_{21}e^{\theta_{\epsilon,1}-\theta_{\epsilon,1}^*}\Omega_{22}
\!-\!a_{11}a_{22}e^{\theta_{\epsilon,1}-\theta_{\epsilon,2}^*}\Omega_{12}
 \!-\!a_{12}a_{21}e^{\theta_{\epsilon,2}-\theta_{\epsilon,1}^*}\Omega_{21}
 \!+\! a_{12}a_{22}e^{\theta_{\epsilon,2}-\theta_{\epsilon,2}^*}\Omega_{11})}
 {\Omega_{12}\Omega_{21}-\Omega_{11}\Omega_{22}},
\ene
where $\Omega_{kj},\, k,j=1,2$ are given by
\begin{align}\no
\Omega_{kj}=\frac{a_{1k}a_{1j}e^{\theta_{\epsilon,k}^{*}+\theta_{\epsilon,j}}
+a_{2k}a_{2j}e^{-\theta_{\epsilon,k}^{*}-\theta_{\epsilon,j}}}{\lambda_k^{*}-\lambda_j}.
\end{align}

\begin{figure}[!t]
    \centering
\vspace{-0.1in}
  {\scalebox{0.36}[0.36]{\includegraphics{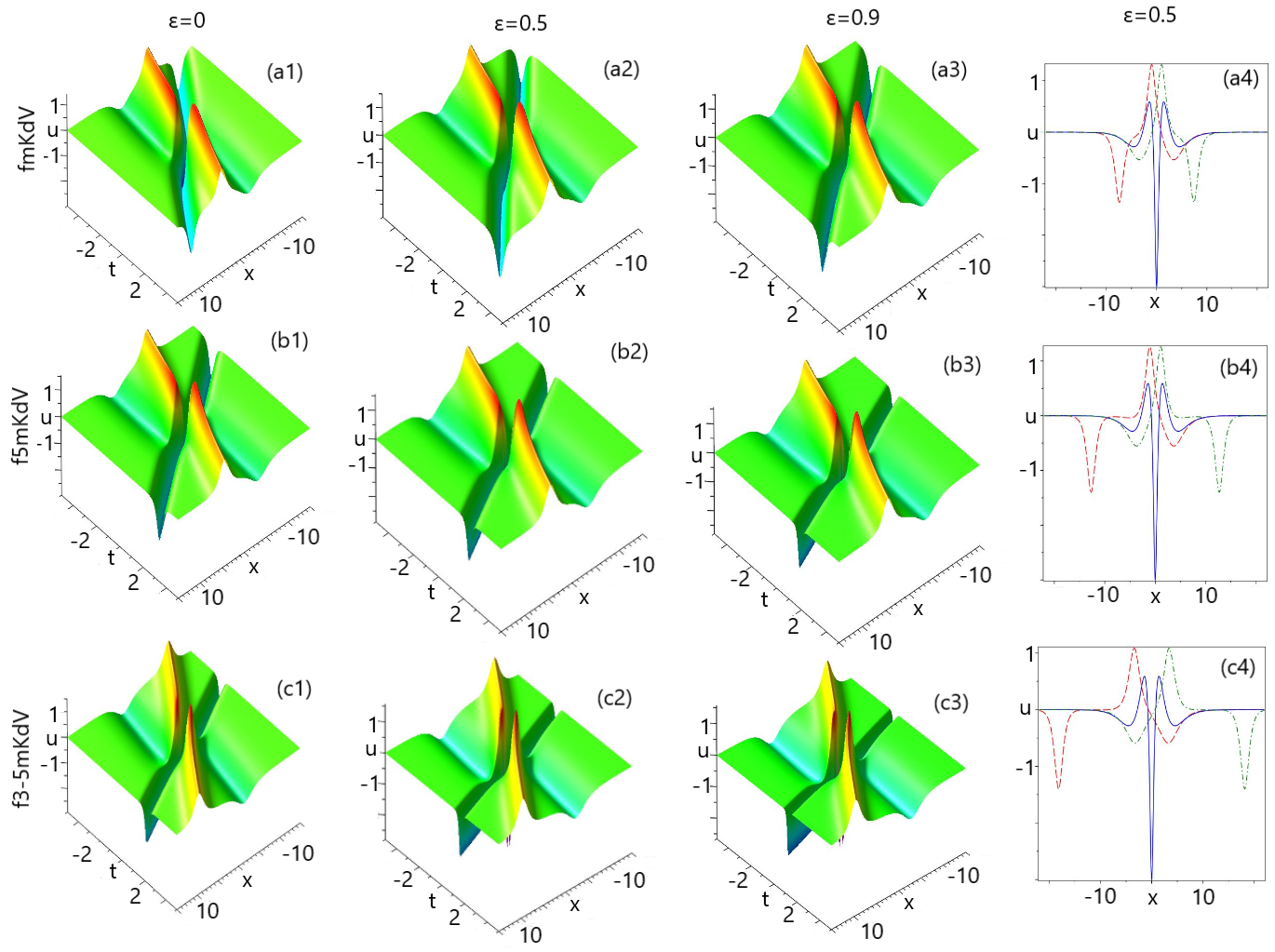}}}
\vspace{-0.05in}
\caption{\small $N=N_1=3.$ Elastic interactions of fractional two-dark-soliton and one-bright-soliton solution for $\lambda_1=0.3i,\, \lambda_2=0.5i,\, \lambda_3=0.7i,\, a_{ij}=1,\,i=1,2; j=1,2,3$. (a1)-(a4) $\alpha_3=1, \alpha_5=0$ (fmKdV equation), (b1)-(b4) $\alpha_3=0,\, \alpha_5=1$ (one W-shaped and one bright solitons of the f5mKdV equation), (c1)-(c4): $\alpha_3=1,\, \alpha_5=1$ (f35mKdV equation). (a4)-(c4) the fractional one-soliton curves with $\epsilon=0.5$ at different times $t=-2$ (dashed line), $t=0$ (solid line) and $t=2$ (dash-dotted line).}
  \label{fig3pure}
\end{figure}

Figures~\ref{fig2pure}(a1)-(c4) illustrate, respectively, the elastic interactions of the fractional two-soliton (bright-dark) solutions of the fmKdV, f5mKdV and f35mKdV equations for two spectral parameters $\lambda_1=0.4i,\, \lambda_2=0.6i$ and different parameter $\epsilon=0,\, 0.5,\, 0.9$, which are all interactions of one left-going travelling-wave  bright soliton and another right-going travelling-wave dark soliton without dissipating or spreading out (see Figs.~\ref{fig2pure}(c1)-(c4)).

\v {\bf Case 3a.}\, As $N=N_1+2N_2=3,\, N_1=N_2=1$, that is, three spectral parameters $\lambda_1\in i\mathbb{R}^+,\, \lambda_2\in\mathbb{C}^{+}\backslash i\mathbb{R}^{+}$ and $\lambda_3=-\lambda_2^*$ are considered,  the fractional two-soliton (one dark soliton and one breather/one W-shaped soliton) solutions of Eq.~(\ref{fmkdv}) are given in the form (\ref{u}).

Figures~\ref{fig4}(a1)-(a4), (b1)-(b4), and (c1)-(c4) illustrate the elastic interactions of the fractional one-breather and one-dark-soliton solutions of the fmKdV equation for three spectral parameters $\lambda_3=0.6i\, \lambda_2=-\lambda_3^*=0.6+0.4i$, the f5mKdV and f35mKdV equations for $\lambda_1=0.7i$,  $\lambda_2=-\lambda_3^*=0.8+0.2i$,
and different parameters $\epsilon=0,\, 0.5,\, 0.9$, respectively, which are all the interactions of one left-going travelling-wave breather and another right-going travelling-wave dark soliton without dissipating or spreading out. However, Figures~\ref{fig4}(d1)-(d4) exhibit the elastic interaction of the fractional one
W-shaped-soliton and one-dark-soliton solutions of the f5mKdV equation for three spectral parameters $\lambda_1=0.3i$,  $\lambda_2=-\lambda_3^*=0.5+0.5i$ and different parameters $\epsilon=0,\, 0.5,\, 0.9$,  which are all the interactions of one left-going travelling-wave W-shaped soliton and another right-going travelling-wave dark soliton without dissipating or spreading out.

\v {\bf Case 3b.}\, As $N=N_1=3$, that is, three spectral parameters $\lambda_1,\,\lambda_2,\, \lambda_3\in i\mathbb{R}^+$ (i.e., pure imaginary spectral parameters) are considered, the fractional three-soliton solutions of Eq.~(\ref{fmkdv}) are given in the form (\ref{u}).

Figures~\ref{fig3pure}(a1)-(c4) illustrate, respectively, the elastic interactions of the fractional three-soliton (two-dark-one-bright) solutions of the fmKdV, f5mKdV and f35mKdV equations for three pure imaginary spectral parameters $\lambda_1=0.3i,\, \lambda_2=0.5i,\, \lambda_1=0.7i$ and different parameter $\epsilon=0,\, 0.5,\, 0.9$, which are all interactions of one right-going travelling-wave bright soliton, one left-going travelling-wave bright soliton, and another right-going travelling-wave dark soliton without dissipating or spreading out (see Figs.~\ref{fig3pure}(c1)-(c4)).

 \v In fact, we can also display the more abundant wave structures of the fractional $N$-solitons as $N>3$.

\section{Conclusions and discussions}

In conclusion, we have studied the integrable combined fractional higher-order mKdV hierarchy with the aid of the Riesz fractional derivative, and their corresponding anomalous dispersion relations. Moreover, the combined fhmKdV hierarchy can be characterized via completeness of squared scalar eigenfunctions. Moreover, the Riemann-Hilbert approach is used to explore
the fractional multi-soliton solutions of combined fhmKdV hierarchy. In particular, we display the fractional one-soliton (one-breather) solutions, interactions of fractional two-soliton (e.g., one breather and one dark soliton, one bright and one dark solitons, one dark and one W-shaped solitons, one bright soliton and two dark solitons) of the fmKdV, f5mKdV, and f35mKdV equations. These found fractional multi-soliton solutions will be useful to understand the related super-dispersion transports of nonlinear waves in fractional nonlinear media. Moreover, the RH approach can also be extended to other integrable fractional nonlinear wave equations.



\v\noindent {\bf Acknowledgements}

This work was supported by the National Natural Science Foundation of China (Grant No.11925108).





%


\begin{thebibliography}{99} \setlength{\itemsep}{-0.8mm}
\makeatletter
\small

\bibitem{fc-book} S. Das, {\it Introduction to Fractional Calculus} (Springer, Berlin, 2011).

\bibitem{fc0} J. A. Tenreiro Machado, A. M. S. F. Galhano, and J. J. Trujillo, On development of fractional calculus during the last
fifty years, Scientometrics 98 (2014) 577-582.

\bibitem{fc1} K. B. Oldham  and J. Spanier, {\it The Fractional Calculus: Theory and Application of Differentiation and
Integration to Arbitrary Order} (Academic Press, New York, 1974).

\bibitem{fc2} R. Hilfer, {\it Applications of Fractional Calculus in Physics} (World Scientific, Singepore, 2000).



\bibitem{frw} B. D. Hughes, E. W. Montroll,  and M. F. Shlesinger, Fractal random walks, J. Stat. Phys. 28 (1982) 111-126.

\bibitem{frw2} R. Metzler and J. Klafter, The random walk’s guide to anomalous diffusion: a fractional dynamics approach,
 Phys. Rep. 339 (2000) 1–77.

 \bibitem{plp} K. Harigaya, M.Ibe, K. Schmitz, and T. T. Yanagida, Dynamical fractional chaotic inflation--Dynamical generation of a fractional power-Law potential for chaotic inflation, Phys. Rev. D 90 (2014) 123524.

\bibitem{31} C. Tadjeran, M. M. Meerschaert,  and H. P. Scheffler, A second-order accurate numerical approximation for the fractional diffusion equation, J. Comput. Phys. 213 (2006) 205-213.

\bibitem{prl09} I. Bronstein, Y. Israel, E. Kepten, S. Mai, Y. Shav-Tal, E. Barkai, and Y. Garini, Transient anomalous diffusion of telomeres in the nucleus of mammalian cells, Phys. Rev. Lett. 103 (2009) 018102.

\bibitem{32} V. V. Anh  and N. N. Leonenko, Harmonic analysis of random fractional diffusion-wave equations, Appl. Math. Comput.
   141 (2003) 77-85.

\bibitem{33} L. R. da Silva, L. S. Lucena, E. K. Lenzi, R. S. Mendes,  and K. S. Fa, Fractional and nonlinear diffusion equation: additional results, Physica A 344 (2004) 671-676.


\bibitem{35} R. Hilfer, Fractional diffusion based on Riemann-Liouville fractional derivatives,  J. Phys. Chem. B 104 (2000) 3914-4917.

\bibitem{27} B. N. Achar, J. W. Hanneken,  and T. Calarke, Damping characteristics of a fractional oscillator, Physica A 339 (2004) 311-319.

\bibitem{28} F. Mainardi, Fractional relaxation-oscillation and fractional diffusion-wave phenomena, Chaos. Soliton. Fract. 7 (1996) 1461-1477.


\bibitem{30} Y. E. Ryabov  and A. Puzenkoa, Damped oscillation in view of the fractional oscillator equation, Phys. Rev. B 66 (2002) 184-201.


\bibitem{41} M. D. Ortigueira, On the initial conditions in continuous-time fractional linear systems, Signal. Process 83 (2003) 2301-2309.

\bibitem{42} R. Metzler  and T. F. Nonnenmacher, Fractional relaxation process and fractional rheological models for the description of a class of viscoelastic matrials, Int. J. Plasticity 19 (2003) 941-959.

\bibitem{43} M. Raberto, E. Scalas,  and F. Maniardi, Waiting-times and returns in high-frequency financial data: an empirical study, Phys. A 314 (2002) 749-755.

\bibitem{44} L. Sabatelli, S. Keating, J. Dudley,  and P. Richmond, Waiting time distributions in financial markets, Eur Phys. J. B 27 (2002) 273-275.


 \bibitem{soliton-2} M. J. Ablowitz and H. Segur, {\em Solitons and the inverse scattering transform}
  (SIAM, Philadelphia, 1981).

  \bibitem{soliton-3} M. J. Ablowitz, B. Prinari, and A. Trubatch, {\em Discrete and continuous nonlinear Schr\"odinger systems} ( Cambridge University Press, Cambridge, 2004).

 \bibitem{soliton} M. J. Ablowitz  and P. A. Clarkson, {\em Soliton, Nonlinear Evolution Equations and Inverse Scattering} (Cambridge Univeristy Press, Cambridge, 1991).


 \bibitem{GGKM} C.~S. Gardner, J.~M. Greene, M.~D. Kruskal,  and R.~M. Miura, Method for solving the
  Korteweg-de Vries equation, Phys. Rev. Lett. 19 (1967) 1095.


\bibitem{carr18}  U. Al Khawaja, M. Al-Refai, G. Shchedrin, and L. D. Carr, High-accuracy power series solutions with arbitrarily large radius of convergence for the fractional nonlinear Schr\"odinger-type equations, J. Phys. 51, 235201 (2018).

\bibitem{boris20} Y. Qiu, B. A. Malomed, D. Mihalache, X. Zhu, X. Peng, and Y. He, Stabilization of single-and multi-peak solitons in the fractional nonlinear Schr\"odinger equation with a trapping potential, Chaos Solitons Fractals 140, 110222 (2020).

\bibitem{boris21} B. A. Malomed, Optical solitons and vortices in fractional media: A mini-review of recent results,
Photonics 8 (2021) 353.

\bibitem{fc-book2} K.S. Miller and B. Ross, {\it An Introduction to the Fractional Calculus and Fractional Differential Equations}
  (John Wiley \& Sons Inc., New York, 1993).

\bibitem{riesz} M. Riesz, L'int\'{e}grable de Riemann-liouville et le probl\'{e}me de Cauchy, Acata Math. 81 (1949) 1-222.

\bibitem{lischke} A. Lischke, G. Pang, M. Gulian, F. Song, C. Glusa, X. Zheng, Z. Mao, W.C ai, M. M. Meerschaert, M. Ainsworth, and G. E. Karniadakis, What is the fractional Laplacian? A comparative review with new results, J. Comput. Phys. 404 (2020) 109009.

\bibitem{fmark} M. J. Ablowitz, J. B. Been, and L. D. Carr, Fractional integrable nonlinear soliton equations, Phys. Rev. Lett. 128 (2022) 184101.


\bibitem{ab22}  M. J. Ablowitz,  J. B. Been, and L. D. Carr,  Integrable fractional modified Korteweg-de Vries, sine-Gordon,
and sinh-Gordon equations, arXiv: 2203.13755.


\bibitem{novikov} S. Novikov, S. Manakov, L. Pitaevskii, and V. Zakharov, {\it Theory of Solitons: the Inverse Scattering Method} (Spring: New York, 1984).

\bibitem{rh1} V. S. Shchesnovich  and I. V. Barashenkov, Soliton-radiation coupling in the parametrically driven, damped nonlinear
Schr\"{o}dinger equation, Phys. D 164 (2002) 83-109.

\bibitem{yang09} J. Yang and D. J. Kaup, Squared eigenfunctions for the Sasa-Satsuma equation,
J. Math. Phys. 50 (2009) 023504.

\bibitem{rh3}D. Wang, D. Zhang,  and J. Yang,  Integrable properties
of the general coupled nonlinear Schr\"{o}dinger equations, J. Math. Phys. 51 (2010) 023510.


\bibitem{guo12} B. Guo and L. Ling, Riemann-Hilbert approach and N-soliton formula for coupled derivative Schr\"odinger equation, J. Math. Phys. 53 (2012) 073506.

\bibitem{fan16} Y. Xiao  and E. Fan, A Riemann–Hilbert approach to the Harry-Dym equation on the line, Chin. Ann. Math. Ser. B 37
(2016) 373–384.



\bibitem{rh6} J. Wu  and X. Geng, Inverse scattering transform and soliton classification of the coupled modified Korteweg-de
Vries equation, Commun. Nonlinear Sci. Numer. Simul. 53, 83-93 (2017).


\bibitem{rh5} W. X. Ma, Riemann-Hilbert problems and N-soliton solutions for a coupled mKdV system, J. Geom. Phys. 132, 45-
54 (2018).


\bibitem{rh2} B. Yang  and Y. Chen, High-order soliton matrices for Sasa-Satsuma equation via local Riemann-Hilbert problem. Nonlinear Anal. Real World Appl. 45 (2019) 918-941.






\bibitem{rh7} W. Peng, S. F. Tian, X. Wang, T. Zhang,  and Y. Fang, Riemann-Hilbert method and multi-soliton solutions for
three-component coupled nonlinear Schr\"{o}dinger equations, J. Geom. Phys. 146 (2019) 103508.




\bibitem{nls} V. E. Zakharov and A. B. Shabat, Exact theory of two-dimensional self-focusing and one-dimensional self-modulation of waves in nonlinear media, Sov. Phys. JETP  34, 62-69 (1972) [Zh. Eksp. Teor. Fiz. 61 (1971) 118-134].

\bibitem{wadati73} M. Wadati, The modified Korteweg-de Vries equation, J. Phys. Soc. Jpn. 34 (1973) 1289-1296.

\bibitem{akns} M. J. Ablowitz, D. J. Kaup, A. C. Newell, and H. Segur, The inverse scattering transform-Fourier analysis for
nonlinear problems, Stud. Appl. Math. 53 (1974) 249-315.


\bibitem{hmkdv1} M. Ito, An extension of nonlinear evolution equations of the KdV (mKdV) type to higher orders,
 J. Phys. Soc. Jpn. 49 (1980) 771.

\bibitem{hmkdv2} Y. Matsuno, Bilinearization of nonlinear evolution equations. II. Higher-order modified Korteweg-de Vries equations, J. Phys. Soc. Jpn. 49 (1980) 787.


\bibitem{kaup76} D. J. Kaup, Closure of the squared Zakharov-Shabat eigenstates, J. Math. Anal. Appl. 54 (1976) 849-864.



\end{thebibliography}
  \end{document}